# Knowledge, awareness, and bisimulation

Hans van Ditmarsch
LORIA, CNRS - Univ. de Lorraine
hvd@us.es

Tim French
University of Western Australia
tim@csse.uwa.edu.au

Fernando R. Velázquez-Quesada
University of Seville
FRVelazquezQuesada@us.es

Yì N. Wáng
Bergen University College
yi.wang@hib.no

## ABSTRACT
We compare different epistemic notions in the presence of awareness of propositional variables: the logics of implicit knowledge (in which explicit knowledge is definable), explicit knowledge, and speculative knowledge. Different notions of bisimulation are suitable for these logics. We provide correspondence between bisimulation and modal equivalence on image-finite models for these logics. The logic of speculative knowledge is equally expressive as the logic of explicit knowledge, and the logic of implicit knowledge is more expressive than both. We also provide axiomatizations for the three logics — only the one for speculative knowledge is novel. Then we move to the study of dynamics by recalling action models incorporating awareness. We show that any conceivable change of knowledge or awareness can be modelled in this setting, we give a complete axiomatization for the dynamic logic of implicit knowledge. The dynamic versions of all three logics are, surprising, equally expressive.

## 1. INTRODUCTION

*Motivating example.* Explicit knowledge is often defined as implicit knowledge plus awareness, with implicit knowledge given by the standard modal box [4, 9]. Thus, to express that 'agent $i$ knows $\varphi$ explicitly', $K_i^E \varphi$, we use formulas of the form $\Box_i \varphi \wedge A_i \varphi$. In such frameworks, awareness is typically modelled as a function $\mathcal{A}$ that indicates the set of formulas each agent is aware of at each state; hence, $A_i \varphi$ is true at state $s$ iff $\varphi \in \mathcal{A}_i(s)$. When the agents' awareness consists of all formulas built from a subset of atoms (the so-called propositional awareness), we can simply associate with a formula $\varphi$ the set of atoms $Q \subseteq P$ occurring in $\varphi$, and we can then say that $A_i \varphi$ is true at state $s$ iff $Q \subseteq \mathcal{A}_i(s)$.

This definition of explicit knowledge can lead to counter-intuitive situations. Consider the following models.

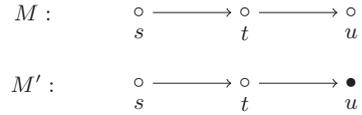

Model $M$ has a domain $\{s, t, u\}$, a single agent $i$ with accessibility relation $R = \{(s, t), (t, u)\}$, atom $p$ true in all states, and the agent is aware of $p$ only in state $s$. Awareness is not depicted. Model $M'$ is like $M$, except that $p$ is now false in $u$ (the black dot).

As mentioned, the agent knows explicitly a given $\varphi$ at a given state iff she is aware of the formula in that state and $\varphi$ is true in all accessible states. Let us apply this to the depicted structures. In both, the agent is unaware of $p$ at state $t$, and therefore of the value of $p$ in $u$: she should see $(M, t)$ and $(M', t)$ as identical, and therefore $(M, s)$ and $(M', s)$ as well. We propose a notion of bisimilarity for which $(M, s)$ and $(M', s)$ are bisimilar.

Now here is the surprise: in the language with awareness and modal box, states $(M, s)$ and $(M', s)$ are not modally equivalent. Given explicit knowledge $K_i^E \varphi$ as $\Box_i \varphi \wedge A_i \varphi$, consider $K_i^E \Box_i p$. This is true in $(M, s)$ but false in $(M', s)$.

In logics of awareness [4] it is common only to consider models for knowledge (equivalence relations) and belief. However, as always in multi-agent logics, it is elementary to transform a single-agent model with directed (asymmetric) accessibility into a multi-agent model where intersecting equivalence classes for agents force such asymmetry. For example, consider the following.

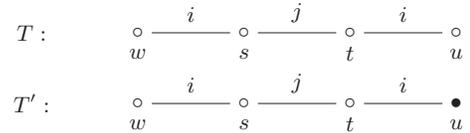

Models $T$ and $T'$ have equivalence accessibility relations (a line represents a two-directions arrow, with reflexive and transitive arrows omitted) for agents $i$ and $j$. Agent $i$ is aware of $p$ in the states $w$, and unaware of $p$ in every other state; agent $j$ is unaware of $p$ in every state. The only difference between $T$ and $T'$ is that $p$ is true at $(T, u)$ and false in $(T', u)$. Again, intuitively, these models are the same from agent $i$'s perspective. But $K_i^E \Box_j \Box_i p$ is true above and false below.

The problem here is the presence of the $\Box$. If the $K^E$ operator is not defined by abbreviation but a primitive in the language, then the models cannot be distinguished, as we will prove. Explicit possibility $L^E$ seems another desirable primitive, as it is not the dual of explicit knowledge (both require awareness). This led us to the comparison of logics where different epistemic notions are primitive. Instead of $K^E$ and $L^E$ as primitives, it turns out that we can equally well take $K^E$ and $A$ (awareness) as primitive, and this language then contrasts nicely with the initial one with $\Box$ and $A$ as primitives. A third epistemic notion is also in our focus: speculative knowledge $K^S$ [20, 21], and with that, the language with $K^S$ and $A$. An agent $i$ speculatively knows $\varphi$, $K_i^S \varphi$, if in any $i$-accessible state, in any state indistinguishable from that as far as awareness of $i$ is concerned, $\varphi$ is true. This is exactly the sense in which $(M,s)$ and $(M',s')$, or $(T,w)$ and $(T',w)$, are similar for $i$.

*Our results.* This paper addresses the question of what a proper notion of knowledge should be in the presence of awareness, and what the proper notion of bisimulation should be in structures encoding knowledge and awareness; how these choices interact; and how adding dynamics of knowledge and awareness further affects this. We present two notions of bisimulation for the Fagin and Halpern structures of [4], *standard bisimulation* and *awareness bisimulation*; and we present three logics, all in the presence of operators $A_i \varphi$ for awareness of variables occurring in $\varphi$: *the logic of implicit knowledge* (with $\Box_i$, so that $K_i^E$ is definable), *the logic of explicit knowledge* (with $K_i^E$), and *the logic of speculative knowledge* (with $K_i^S$), summarily introduced above as knowledge modulo speculation over unaware variables. We then show that, on image-finite models, standard bisimilarity corresponds to modal equivalence in the logic of implicit knowledge, but that awareness bisimilarity corresponds to modal equivalence in the logic of explicit knowledge, and also to modal equivalence in the logic of speculative knowledge. We continue by listing various expressivity results, mainly that the logic of implicit knowledge is (strictly) more expressive than the logic of explicit knowledge (reminiscent of [9]). After that we give axiomatizations for our three logics. The logic of implicit knowledge was already axiomatized in [4] and the logic of explicit knowledge in [9], but the axiomatization for the logic of speculative knowledge is novel. Then we investigate the dynamics of awareness and of knowledge, by way of *epistemic awareness action models*. The dynamic logic of speculative knowledge has already been reported in [22]. Here, we show that on the class of finite models every conceivable change of (implicit, explicit, or speculative) knowledge or awareness can be modelled in an epistemic awareness action model. Finally, we give a complete axiomatization for the dynamic logic of implicit knowledge. The dynamic versions of the logics are, surprising, equally expressive. This also gives us the axiomatization for the dynamic logic of explicit knowledge.

*Overview of the literature.* Our work is rooted in the tradition of epistemic logic [13] and in particular multi-agent epistemic logic [15, 5], in various works on the interaction between awareness and knowledge [4, 16, 9, 11, 12, 8, 10], and in modal logical research in propositional quantification, starting with [6] and followed up by work on bisimulation quantifiers [24, 14, 7].

Works treating awareness either follow a more *semantically* flavoured approach, where awareness is defined in terms of a set of propositional variables [17, 11], or a more *syntactically* flavoured approach, where awareness concerns all formulas of the language in a given set, in order to model 'limited rationality' of agents [4, 19]. Our proposal falls straight into the semantic corner: within the limits of their awareness, agents are fully rational.

## 2. LOGICS FOR AWARENESS

Throughout the contribution, given are a countable non-empty set of atomic propositions $P$ and a (disjoint) finite non-empty set of agents $N$.

**Definition 1 (Epistemic awareness model)** *An epistemic awareness model is a tuple $M = (S, R, \mathcal{A}, V)$ where*

- *$S$ (also denoted by $\mathcal{D}(M)$) is a non-empty set of* states;
- *$R : N \to \mathcal{P}(S \times S)$ is an* accessibility *function;*
- *$\mathcal{A} : N \to S \to \mathcal{P}(P)$ is an* awareness *function;*
- *$V : P \to \mathcal{P}(S)$ is a* valuation.

*A pair $(M, s)$ with $s \in S$ is an* epistemic awareness state.

We write $R_i$ for $R(i)$, $\mathcal{A}_i$ for $\mathcal{A}(i)$, and $R_i(s)$ for $\{t \in S \mid R_i(s,t)\}$. An epistemic awareness model is *image-finite* if all $R_i(s)$ are finite.

An epistemic awareness model is simply an epistemic model plus a propositional awareness function. We associate two notions of bisimulation [18, 3] with this. Standard bisimulation is the more obvious one, but awareness bisimulation is evidently the more suitable notion in view of our introductory examples. The motivation for awareness bisimulation was the lattice of state spaces in [11]; see [20, 21] for details.

**Definition 2 (Standard bisimulation)** *Let $Q \subseteq P$. A $Q$ standard bisimulation* between epistemic awareness models $M = (S, R, \mathcal{A}, V)$ and $M' = (S', R', \mathcal{A}', V')$ is a relation $\mathfrak{R}[Q] \subseteq (S \times S')$ such that, for every $(s, s') \in \mathfrak{R}[Q]$, for every agent $i \in N$, and for every $p \in Q$:

- **atoms:** $s \in V(p)$ iff $s' \in V'(p)$;
- **aware:** $Q \cap \mathcal{A}_i(s) = Q \cap \mathcal{A}'_i(s')$;
- **forth:** *if $t \in R_i(s)$ then there is a $t' \in R'_i(s')$ such that $(t, t') \in \mathfrak{R}[Q]$;*
- **back:** *if $t' \in R'_i(s')$ then there is a $t \in R_i(s)$ such that $(t, t') \in \mathfrak{R}[Q]$.*

$(M, s)$ and $(M', s')$ are $Q$ standard bisimilar, notation $(M, s) \simeq_Q (M', s')$, if there is a $Q$ standard bisimulation between $M$ and $M'$ that contains $(s, s')$.

**Definition 3 (Awareness bisimulation)** *As Definition 2 but with the following clauses for* **forth** *and* **back** *instead.*

- **forth:** *if $t \in R_i(s)$ then there is a $t' \in R'_i(s')$ such that $(t, t') \in \mathfrak{R}[Q \cap \mathcal{A}_i(s)]$;*
- **back:** *if $t' \in R'_i(s')$ then there is a $t \in R_i(s)$ such that $(t, t') \in \mathfrak{R}[Q \cap \mathcal{A}'_i(s')]$.*

*where $\mathfrak{R}[Q \cap \mathcal{A}_i(s)]$ is a $Q \cap \mathcal{A}_i(s)$ awareness bisimulation and $\mathfrak{R}[Q \cap \mathcal{A}'_i(s')]$ is a $Q \cap \mathcal{A}'_i(s')$ awareness bisimulation. The notation for $Q$ awareness bisimilarity is $(M, s) \underline{\leftrightarrow}_Q (M', s')$.*

In an awareness bisimulation, the perspective of the agent is restricted to the variables that she is aware of, therefore in the **back** and **forth** steps bisimulation is only checked for the variables in $Q \cap \mathcal{A}_i(s)$ instead of the variables in $Q$. The following is therefore obvious.

**Proposition 4** Let $(M, s)$ and $(M', s')$ be epistemic awareness models, and $Q \subseteq P$. If $(M, s) \simeq_Q (M', s')$, then $(M, s) \underline{\leftrightarrow}_Q (M', s')$.

**Example 1** Awareness bisimularity does not imply standard bisimularity. The epistemic awareness states $(M, s)$ and $(M', s)$ of the introduction are $\{p\}$ awareness bisimilar. To see this, observe that $(M, u)$ and $(M', u')$ are $\emptyset$ awareness bisimilar; then, because of this, because $\{p\} \cap \mathcal{A}_i(t) = \emptyset$ and because $t$'s states coincide in $p$'s truth value and in $i$'s awareness of $p$, epistemic awareness states $(M, t)$ and $(M', t')$ are $\{p\}$ awareness bisimilar. In turn, this, the fact that $\{p\} \cap \mathcal{A}_i(s) = \{p\}$ and the fact that $s$ and $s'$ coincide in $p$'s truth value and in $i$'s awareness of $p$, make epistemic awareness states $(M, s)$ and $(M', s')$ $\{p\}$ awareness bisimilar too.

However, $(M, s)$ and $(M', s)$ are not $\{p\}$ standard bisimilar because, in turn, $(M, t)$ and $(M', t)$ are not $\{p\}$ standard bisimilar, and this is because $(M, u)$ and $(M', u)$ are not $\{p\}$ standard bisimilar: they differ in $p$'s truth-value.

**Definition 5 (Language)** The language $\mathcal{L}(\Box, K^E, L^E, K^S, A)$ is defined as follows, where $p \in P$ and $i \in N$.

$$\varphi ::= \top \mid p \mid \neg\varphi \mid \varphi \wedge \varphi \mid \Box_i \varphi \mid K_i^E \varphi \mid L_i^E \varphi \mid K_i^S \varphi \mid A_i \varphi$$

Given a language $\mathcal{L}$, $\mathcal{L}|Q$ is the language with the propositional variables restricted to $Q \subseteq P$.

We typically consider languages for subsets of these inductive rules. We write $\mathcal{L}^\Box$ for $\mathcal{L}(\Box, A)$, $\mathcal{L}^E$ for $\mathcal{L}(K^E, A)$, and $\mathcal{L}^S$ for $\mathcal{L}(K^S, A)$, as these three languages are the main focus of our investigations. We assume familiarity with the meaning of propositional constructs, the modal box, and awareness. Implication $\rightarrow$, disjunction $\vee$, equivalence $\leftrightarrow$, and the modal diamond $\diamondsuit_i$ are defined by abbreviation, as usual. Formula $\Box_i \varphi$ sometimes stands for 'the agent implicitly knows $\varphi$', but we also view it as a mere technical background notion. Formula $K_i^E \varphi$ stands for 'the agent *explicitly knows* that $\varphi$', $L_i^E \varphi$ stands for 'the agent *explicitly considers possible* that $\varphi$'. (Explicit knowledge is not the dual of explicit possibility, as both require awareness.) Formula $K_i^S \varphi$ stands for 'the agent *speculatively knows* that $\varphi$'. Speculative possibility $L_i^S \varphi$ is the dual of speculative knowledge and by abbreviation defined as $L_i^S \varphi$ iff $\neg K_i^S \neg \varphi$. More explanations will be given with the semantics.

**Definition 6 (Free variables)** The free variables of a formula $\varphi$ are defined by $v(p) := \{p\}$, $v(\neg\varphi) := v(\varphi)$, $v(\varphi \wedge \psi) := v(\varphi) \cup v(\psi)$ and $v(Y\varphi) := v(\varphi)$, where $Y$ is one of $\Box_i, A_i, K_i^E, L_i^E, K_i^S$.

**Definition 7 (Semantics)** Let $(M, s)$ be an epistemic awareness state, with $M = (S, R, \mathcal{A}, V)$. The non-propositional clauses are

$(M, s) \models \Box_i \varphi$ iff $\forall t \in R_i(s), (M, t) \models \varphi$
$(M, s) \models A_i \varphi$ iff $v(\varphi) \subseteq \mathcal{A}_i(s)$
$(M, s) \models K_i^E \varphi$ iff $v(\varphi) \subseteq \mathcal{A}_i(s)$ and $\forall t \in R_i(s), (M, t) \models \varphi$
$(M, s) \models L_i^E \varphi$ iff $v(\varphi) \subseteq \mathcal{A}_i(s)$ and $\exists t \in R_i(s), (M, t) \models \varphi$
$(M, s) \models K_i^S \varphi$ iff $\forall t \in R_i(s), \forall (M', t') \underline{\leftrightarrow}_{\mathcal{A}_i(s)} (M, t), (M', t') \models \varphi$

Model validity $M \models \varphi$ and validity $\models \varphi$ are defined as usual. The logic (i.e., the set of validities) of language $\mathcal{L}^\Box$ is called $\mathsf{L}^\Box$, the logic of $\mathcal{L}^E$ is $\mathsf{L}^E$, and the logic of $\mathcal{L}^S$ is $\mathsf{L}^S$.

We will refer to our standard logics as follows:

- $\mathsf{L}^\Box$: the logic of implicit knowledge
- $\mathsf{L}^E$: the logic of explicit knowledge
- $\mathsf{L}^S$: the logic of speculative knowledge

We pay attention to semantic relations between the non-propositional primitives in Section 5. E.g., it is the case that $A_i \varphi \leftrightarrow (K_i^E \varphi \vee L_i^E \neg \varphi)$.

Speculative knowledge is defined in terms of awareness bisimulation: agent $i$ knows speculatively $\varphi$ at $(M, s)$ iff $\varphi$ is the case in every epistemic awareness state that is $\mathcal{A}_i(s)$ awareness bisimilar to some state $t$ accessible from $s$ in $M$.

Speculative and explicit knowledge are different. For example, any agent knows $p \vee \neg p$ speculatively, even if she is unaware of $p$, because in every possible state $p \vee \neg p$ is true. Nevertheless, the agent only knows $p \vee \neg p$ explicitly when she is aware of $p$.

Speculative and implicit knowledge are also different. The agent may implicitly know $p$, but she cannot speculatively know that, because she can speculate about $p$ being false. And if $p$ were false, she cannot know that $p$.

More convincing examples of speculative knowledge involve dynamics. Suppose that the agent explicitly knows $q$ but is unaware of $p$. She then speculatively knows: "If $p$ is false then even if I were to become aware of $p$ I cannot explicitly know that $p$ and $q$ are both true." (In the extended logic of Section 7 this is formally $\neg p \to [\mathsf{A}^{+p}] \neg K_i^E (p \wedge q)$, where $[\mathsf{A}^{+p}]$ is a dynamic modal operator.) But she does not explicitly know that, because she is unaware of $p$, and $p$ occurs in the formula. For more intuitions, see [20, 21, 22].

**Definition 8 (Modal equivalence)** Awareness epistemic states $(M, s)$ and $(M', s')$ are modally equivalent in a language $\mathcal{L}$ up to $Q \subseteq P$, notation $(M, s) \equiv_Q^\mathcal{L} (M', s')$, if for all $\varphi \in \mathcal{L}|Q$, $(M, s) \models \varphi$ iff $(M', s') \models \varphi$. For $\mathcal{L} = \mathcal{L}^\Box, \mathcal{L}^E, \mathcal{L}^S$ we write for that, respectively, $\equiv_Q^\Box$, $\equiv_Q^E$, and $\equiv_Q^S$.

**Example 2** Consider the first introductory example. The formula $K_i^E \Box_i p$ is true in $(M, s)$ and false in $(M', s)$. The models are not modally equivalent in the logic $\mathcal{L}^\Box$. But they are modally equivalent in the logic $\mathcal{L}^E$ (without $\Box$), as we will show later.

## 3. BISIMILARITY AND MODAL EQUIVALENCE

For the logic of implicit knowledge we have, as expected, that standard bisimilarity implies modal equivalence in $\mathcal{L}^\Box$. Moreover, in the class of image-finite models, modal equivalence in $\mathcal{L}^\Box$ implies standard bisimilarity. (Let $(M, s)$ and $(M', s')$ be epistemic awareness models, and $Q \subseteq P$...)

**Proposition 9**
$(M, s) \simeq_Q (M', s')$ implies $(M, s) \equiv_Q^\Box (M', s')$.

PROOF. The proof is standard, by induction on $\varphi$. The case for formulas of the form $A_i \varphi$ follows from the **aware** clause in Definition 2.

**Proposition 10** On image-finite models:
$(M, s) \equiv_Q^\Box (M', s')$ implies $(M, s) \simeq_Q (M', s')$.

PROOF. Again, the proof is standard. For proving the **aware** clause, we use modal equivalence with respect to formulas of the form $A_i \varphi$.

**Theorem 11** *On image-finite models:*
$(M, s) \simeq_Q (M', s')$ *iff* $(M, s) \equiv_Q^\square (M', s')$.

For the logic of explicit knowledge the correspondence is between awareness bisimulation and modal equivalence in $\mathcal{L}^E$. We recall that awareness bisimulation is a weaker notion than standard bisimulation.

**Proposition 12**
$(M, s) \underline{\leftrightarrow}_Q (M', s')$ *implies* $(M, s) \equiv_Q^E (M', s')$.

PROOF. We show the following:

> Let $\varphi \in \mathcal{L}^E$ and let $Q \subseteq P$ such that $v(\varphi) \subseteq Q$. Then for any $(M, s)$ and $(M', s')$, $(M, s) \underline{\leftrightarrow}_Q (M', s')$ implies that: $(M, s) \models \varphi$ iff $(M', s') \models \varphi$.

In this formulation it is important that $\varphi$ is chosen before $Q$, and both $\varphi$ and $Q$ before the models, so that the inductive hypothesis may be used on a subformula of $\varphi$ for *another* subset of $P$ than the initial $Q$ (and for any models). Again, the proof goes by induction on $\varphi$; all cases are trivial except $K_i^E \varphi$.

**Case** $K_i^E$  Assume $(M, s) \underline{\leftrightarrow}_Q (M', s')$, and suppose that $(M, s) \models K_i^E \varphi$ with $v(K_i^E \varphi) \subseteq Q$ (which implies $v(\varphi) \subseteq Q$). By semantic interpretation, $v(\varphi) \subseteq \mathcal{A}_i(s)$ and every state in $R_i(s)$ satisfies $\varphi$. First, take any $t' \in R_i'(s')$. By **back** there is a $t \in R_i(s)$ such that $(M, t) \underline{\leftrightarrow}_{Q \cap \mathcal{A}_i(s)} (M', t')$. But $t \in R_i(s)$ so $(M, t) \models \varphi$. Moreover, $v(\varphi) \subseteq Q$ and $v(\varphi) \subseteq \mathcal{A}_i(s)$ so $v(\varphi) \subseteq Q \cap \mathcal{A}_i(s)$, and then we can use induction hypothesis to get $(M', t') \models \varphi$. Thus, every element of $R_i'(s')$ satisfies $\varphi$. Second, $(M, s) \underline{\leftrightarrow}_Q (M', s')$ implies $Q \cap \mathcal{A}_i(s) = Q \cap \mathcal{A}_i'(s')$, so from $v(\varphi) \subseteq Q \cap \mathcal{A}_i(s)$ we get $v(\varphi) \subseteq Q \cap \mathcal{A}_i'(s')$ and thus $v(\varphi) \subseteq \mathcal{A}_i'(s')$. Hence, from the two parts we get $(M', s') \models K_i^E \varphi$, as needed. The other direction is similar.

**Proposition 13** *On image-finite models:*
$(M, s) \equiv_Q^E (M', s')$ *implies* $(M, s) \underline{\leftrightarrow}_Q (M', s')$.

PROOF. We will show that the relation of modal equivalence in $\mathcal{L}^E$ with formulas built from atoms in $Q$ is a $Q$ awareness bisimulation, i.e., that $\equiv_Q^E$ satisfies Definition 3.

Suppose that $(M, s) \equiv_Q^E (M', s')$.

- **Atoms**. Take any $p \in Q$ and suppose $s \in V(p)$; then $p \in \mathcal{L}^E|Q$ and $(M, s) \models p$ so $(M', s') \models p$, that is, $s' \in V'(p)$. The other direction is similar.
- **Aware**. Take any $i \in N$, and suppose $p \in Q \cap \mathcal{A}_i(s)$; then $p \in Q$ and $p \in \mathcal{A}_i(s)$. From the latter we get $(M, s) \models A_i p$ and therefore $(M', s') \models A_i p$, that is, $p \in \mathcal{A}_i'(s')$. We already had $p \in Q$, so $p \in Q \cap \mathcal{A}_i'(s')$. The other direction is similar.
- **Forth**. Take any $i \in N$, and suppose $t \in R_i(s)$; we want to find a $t' \in R_i'(s')$ such that $(M, t) \equiv_{Q \cap \mathcal{A}_i(s)}^E (M', t')$. We proceed by contradiction, so suppose no element of $R_i'(s')$ is modally equivalent to $t$ with respect to formulas in $\mathcal{L}^E|(Q \cap \mathcal{A}_i(s))$.
  Observe how $R_i'(s')$ is a finite non-empty set: finite because of image-finiteness, and non-empty because $R_i(s) \neq \emptyset$ iff $(M, s) \models L_i \top$, and since $L_i \top \in \mathcal{L}^E|Q$, we have $(M', s') \models L_i \top$ too. Now, since no element of $R_i'(s')$ is modally equivalent to $t$ with respect to formulas with atoms in $Q \cap \mathcal{A}_i(s)$, then for each $t_k' \in R_i'(s')$ there should be a formula $\varphi_k \in \mathcal{L}^E|(Q \cap \mathcal{A}_i(s))$ that holds at $t$ but fails at $t_k'$.

Now define $\varphi := \varphi_1 \wedge \ldots \wedge \varphi_n$ (with $n$ the cardinality of $R_i'(s')$). We have $(M, t) \models \varphi$ because every $\varphi_k$ is true at $t$, but also $(M', t_k') \not\models \varphi$ for every $k$ because each $\varphi_k$ fails in at least $t_k'$. Moreover, since $\varphi_k \in \mathcal{L}^E|(Q \cap \mathcal{A}_i(s))$ for every $k$, we have $\varphi \in \mathcal{L}^E|(Q \cap \mathcal{A}_i(s))$, and hence $v(\varphi) \subseteq Q \cap \mathcal{A}_i(s)$, that is, $v(\varphi) \subseteq \mathcal{A}_i(s)$. Now, from $t \in R_i(s)$, $(M, t) \models \varphi$ and $v(\varphi) \subseteq \mathcal{A}_i(s)$ we get $(M, s) \models L_i \varphi$. But $(M, s) \not\models L_i \varphi$ because no successor of $s'$ satisfies $\varphi$. Then, $L_i \varphi$ distinguishes between $s$ and $s'$. But since $\varphi \in \mathcal{L}^E|(Q \cap \mathcal{A}_i(s))$, we have $L_i \varphi \in \mathcal{L}^E|(Q \cap \mathcal{A}_i(s))$ and hence $L_i \varphi \in \mathcal{L}^E|Q$: this contradicts $(M, s) \equiv_Q^E (M', s')$. Hence, there should be a state $t' \in R_i'(s')$ such that $(M, t) \equiv_{Q \cap \mathcal{A}_i(s)}^E (M', t')$.
- **Back**. Similar to the **forth** clause.

**Theorem 14** *On image-finite models:*
$(M, s) \underline{\leftrightarrow}_Q (M', s')$ *iff* $(M, s) \equiv_Q^E (M', s')$.

**Example 3** *The formula $K_i^E \square_i p$ distinguishing the models in the introduction is in $\mathcal{L}^\square$ (it is an abbreviation of $\mathcal{A}_i \square_i p \wedge \square_i \square_i p$), but it is not in $\mathcal{L}^E$. It was unclear until now that it does not have an $\mathcal{L}^E$ equivalent. Now it is clear: the models $(M, s)$ and $(M', s')$ are $p$ awareness bisimilar, and therefore modally equivalent in $\mathcal{L}^E$.*

That the language $\mathcal{L}^\square$ of implicit knowledge is aligned with standard bisimulation rather than awareness bisimulation can be seen as a strong argument against the use of this language to specify interactions in epistemic awareness models: it is too rich from the point of view of an agent reasoning about its knowledge and awareness. The language of explicit knowledge $\mathcal{L}^E$ can be seen as its 'explicit' counterpart. Without the aspect of awareness, $\mathcal{L}^\square$ is nothing but the standard multiagent epistemic language, built from the propositional connectives plus operators to talk about what the agent knows and considers possible. Similarly, language $\mathcal{L}^E$ can be seen as (relative to an expressivity result proved in Section 5) built from propositional connectives plus operators to talk about what the agent explicitly knows and explicitly considers possible.

Finally, speculative knowledge. Interestingly, modal equivalence in $\mathcal{L}^S$ for the logic of speculative knowledge is also characterized (on image-finite models) by awareness bisimulation.

**Proposition 15**
$(M, s) \underline{\leftrightarrow}_Q (M', s')$ *implies* $(M, s) \equiv_Q^S (M', s')$.

PROOF. See [21, 22].

**Proposition 16** *On image-finite models:*
$(M, s) \equiv_Q^S (M', s')$ *implies* $(M, s) \underline{\leftrightarrow}_Q (M', s')$.

PROOF. Assume $(M, s) \equiv_Q^S (M', s')$; we will show that the relation $\equiv_Q^S$ defines a $Q$ awareness bisimulation linking $(M, s)$ and $(M', s')$. Clauses **atoms** and **aware** are straightforward; **back** is similar to **forth**.

- **Forth**. We proceed as in Proposition 13. Assume that $t \in R_i(s)$, that the $i$-successors of $t$ are $t_1', \ldots, t_m'$ (a finite number), and that none of those is $Q \cap \mathcal{A}_i(s)$ modally equivalent to $t$. Therefore there are difference formulas $\varphi_1, \ldots, \varphi_m \in \mathcal{L}^S|(Q \cap \mathcal{A}_i(s))$ that are false in $t_1', \ldots, t_m'$, respectively, and true in $t$, so that their conjunction $\psi = \bigwedge_{1..m} \varphi_i$ is true in $t$. This conjunction $\psi$ is also in $\mathcal{L}^S|(Q \cap \mathcal{A}_i(s))$. We now have that $(M, s) \models L_i^S \psi$, as there is an $i$-accessible state from $s$, namely $t$, and a $Q\cap\mathcal{A}_i(s)$ awareness bisimilar state equivalent to $(M,t)$, namely $(M,t)$ itself, such that $(M,t) \models \psi$. (From $(M,t) \equiv_{Q\cap\mathcal{A}_i(s)} (M,t)$ follows by the definition of the bisimulation that $(t,t) \in \mathfrak{R}[Q \cap \mathcal{A}_i(s)]$.) On the other hand, $L_i^S \psi$ is false in $s'$: clearly, $\psi$ is false in any of the states $t'_1, \dots, t'_m$ accessible from $s'$, but any $Q\cap\mathcal{A}_i(s)$ modally equivalent state should also not satisfy $\psi$, as $\psi \in \mathcal{L}^S|(Q \cap \mathcal{A}_i(s))$.

**Theorem 17** *On image-finite models:*
$(M,s)\underline{\leftrightarrow}_Q(M',s')$ *iff* $(M,s) \equiv_Q^S (M',s')$.

## 4. HAVING THE SAME KNOWLEDGE

We can now harvest the benefits from the previous section. We want to characterize when two epistemic awareness states are the same 'from the perspective of an agent', that is, when the agent's knowledge and ignorance is the same in both. This is weaker than being modally equivalent: two epistemic awareness states $(M,s)$ and $(M',s')$ that differ only in a propositional variable $p$ look the same for an agent that is not aware of $p$ in both, and they also look the same for an agent that is aware of $p$ in both but such that the actual state is not accessible. The results for implicit, explicit and speculative knowledge are similar.

**Definition 18 (Same knowledge)** *Let $Q \subseteq P$, and $N' \subseteq N$. Assume epistemic awareness states $(M,s)$ and $(M',s')$.*

- $(M,s)$ *and* $(M',s')$ *describe the same implicit knowledge up to $Q$ for the agents in $N'$ iff, for every agent $i \in N'$ and every formula $\varphi \in \mathcal{L}^\square|Q$, $(M,s) \models \square_i\varphi$ iff $(M',s') \models \square_i\varphi$.*
- $(M,s)$ *and* $(M',s')$ *describe the same explicit knowledge up to $Q$ for the agents in $N'$ iff, for every agent $i \in N'$ and every formula $\varphi \in \mathcal{L}^E|Q$, $(M,s) \models K_i^E\varphi$ iff $(M',s') \models K_i^E\varphi$, and $(M,s) \models L_i^E\varphi$ iff $(M',s') \models L_i^E\varphi$.*
- $(M,s)$ *and* $(M',s')$ *describe the same speculative knowledge up to $Q$ for the agents in $N'$ iff, for every agent $i \in N$ and every formula $\varphi \in \mathcal{L}^S$, $(M,s) \models K_i^S\varphi$ iff $(M',s') \models K_i^S\varphi$.*

*If $(M,s)$ and $(M',s')$ describe the same explicit knowledge for agent $i$ up to (at least) $\mathcal{A}_i(s)$, and $\mathcal{A}_i(s) = \mathcal{A}'_i(s')$, then we can simply say that they describe the same explicit knowledge for agent $i$; and similarly for speculative knowledge.*

To define the same explicit knowledge, we need to refer to both $K^E$ and $L^E$ in the definition (both require awareness). For implicit knowledge and for speculative knowledge the part for the dual diamond version is simply the contraposition of the part for the box version. The 'at least' bit in the final part of the definition is there, because agent $i$ does not explicitly know any formula with variables in $Q \setminus \mathcal{A}_i(s)$, both in $s$ and $s'$.

Write $(M,s)\underline{\leftrightarrow}^i(M',s')$ whenever $(M,s)\underline{\leftrightarrow}_{\mathcal{A}_i(s)}(M',s')$ except for the valuation of atoms in $s$ and $s'$ (i.e., skip clause **atoms** in the root), and except for **back** and **forth** for all other agents than $i$, in the root. Then this $\underline{\leftrightarrow}^i$ equivalence class encodes exactly 'what agent $i$ knows in state $s$'. This works both for explicit knowledge and for speculative knowledge (for implicit knowledge we would require standard bisimulation, but we consider that case of lesser interest).

**Proposition 19** *Let $(M,s)$ and $(M',s')$ be image-finite epistemic awareness models, and $i \in N$. Then $(M,s)\underline{\leftrightarrow}^i(M',s')$ iff $(M,s)$ and $(M',s')$ describe the same explicit / speculative knowledge for agent $i$.*

PROOF. Directly from Theorem 14, resp., Theorem 17.

This structural characterization of explicit knowledge and speculative knowledge, for a given agent, was an important motivation for our investigation.

## 5. EXPRESSIVITY

Two models $(M,s)$ and $(M',s')$ can be distinguished in language $\mathcal{L}$ of logic $\mathsf{L}$ if there is formula $\varphi \in \mathcal{L}$ that is false in $(M,s)$ and true in $(M',s')$; $\varphi$ is called a *distinguishing formula*. A logic $\mathsf{L}$ with language $\mathcal{L}$ is at least as expressive as $\mathsf{L}'$ with language $\mathcal{L}'$ if all pairs of models distinguishable in $\mathcal{L}'$ are also distinguishable in $\mathcal{L}$. A standard way to prove this, is to show that any formula in $\mathcal{L}'$ is equivalent to a formula in $\mathcal{L}$ (and a trivial case is when $\mathcal{L}' \subseteq \mathcal{L}$), and a standard way to disprove it is to show that some pair of models distinguishable in $\mathcal{L}'$ is indistinguishable in $\mathcal{L}$. A logic $\mathsf{L}$ is (strictly) more expressive than a language $\mathsf{L}'$, given a class of models, if $\mathsf{L}$ is at least as expressive as $\mathsf{L}'$ but not vice versa. Instead of expressivity of logics one sometimes talks about the expressivity of languages. The latter is then, of course, relative to a semantics, i.e., it concerns after all a logic.

The expressivity hierarchy is a partial order $<$. We are interested in the relative expressivity of our main logics $\mathsf{L}^\square$, $\mathsf{L}^E$, and $\mathsf{L}^S$. This is a total order: $\mathsf{L}^\square > \mathsf{L}^E = \mathsf{L}^S$. Both terms in the equation are of interest. For example, $\mathsf{L}^E$ and $\mathsf{L}^S$ could just as well have been incomparable. Of further interest is that a number of other logics are equally expressive as $\mathsf{L}^\square$. As we have a good naming device for languages but not for logics we will henceforth **in this section** talk about expressivity of languages, not logics, and we will write all languages in full, e.g., $\mathcal{L}(\square, A)$ instead of $\mathcal{L}^\square$, etc.

**Proposition 20 (Equivalence class of $\mathsf{L}^\square$)**
*The languages $\mathcal{L}(\square, A)$, $\mathcal{L}(\square, K^E)$, $\mathcal{L}(\square, K^E, A)$ and $\mathcal{L}(\square, K^E, L^E)$ are equally expressive.*

PROOF. This follows from the following equivalences:

$$\begin{aligned} K_i^E \varphi &\Leftrightarrow \square_i\varphi \wedge A_i\varphi \\ L_i^E \varphi &\Leftrightarrow \diamondsuit_i\varphi \wedge A_i\varphi \\ A_i\varphi &\Leftrightarrow K_i^E\varphi \vee L_i^E\neg\varphi \quad\Leftrightarrow\quad K_i^E(\varphi \vee \neg\varphi) \end{aligned}$$

**Proposition 21 (Equivalence class of $\mathsf{L}^E$)**
*The languages $\mathcal{L}(K^E, L^E)$, $\mathcal{L}(K^E, A)$, $\mathcal{L}(K^E)$ and $\mathcal{L}(L^E, A)$ are equally expressive.*

PROOF. This follows from the following equivalences:

$$\begin{aligned} K_i^E \varphi &\Leftrightarrow \neg L_i^E\neg\varphi \wedge A_i\varphi \\ L_i^E \varphi &\Leftrightarrow \neg K_i^E\neg\varphi \wedge A_i\varphi \\ A_i\varphi &\Leftrightarrow K_i^E\varphi \vee L_i^E\neg\varphi \quad\Leftrightarrow\quad K_i^E(\varphi \vee \neg\varphi) \end{aligned}$$

**Proposition 22 ($\mathsf{L}^\square > \mathsf{L}^E$)**
$\mathcal{L}(\square, A)$ *is more expressive than* $\mathcal{L}(K^E, A)$.

PROOF. Consider the models $(M,s)$ and $(M,s')$ of the first introductory example. We have seen that they are $\{p\}$ awareness bisimilar, and thus by Proposition 12 modally equivalent in $\mathcal{L}(K^E, A)$. On the other hand, $K_i^E \square_i p \in \mathcal{L}(\square, A)$ distinguishes between the two models. Hence, $\mathcal{L}(\square, A)$ is more expressive than $\mathcal{L}(K^E, A)$.

**Proposition 23 (Equivalence class of $\mathbf{L}^E$, continued)**
$\mathcal{L}(K^S, A)$ and $\mathcal{L}(K^E, A)$ are equally expressive.

PROOF. To show that $\mathcal{L}(K^S, A)$ is at least as expressive $\mathcal{L}(K^E, A)$ it is enough to show that $K^E$ and $L^E$ are expressible in $\mathcal{L}(K^S, A)$. The following obvious (recursive) definitions are sufficient for this.

$$(K_i^E \varphi)' \stackrel{def}{\equiv} K_i^S \varphi' \wedge A_i \varphi \qquad (L_i^E \varphi)' \stackrel{def}{\equiv} L_i^S \varphi' \wedge A_i \varphi$$

For the converse, to show that $\mathcal{L}(K^E, A)$ is at least as expressive as $\mathcal{L}(K^S, A)$, we require the concept of a uniform interpolant [24]. It has been shown that the modal logic $K$ has the uniform interpolation property, that is, if there is a formula $\varphi$ whose variables are taken from the union of the disjoint sets of atoms $Q$ and $R$, then there is a single formula $\varphi^Q$ such that

1. $\varphi \to \varphi^Q$ is valid.
2. the validity of $\varphi \to \gamma$ implies the validity of $\varphi^Q \to \gamma$ for all formulae $\gamma$ not containing any atoms from $R$.

This allows us to define a recursive translation (relative to the set $Q$ of propositional atoms the agent is aware of):

$$(K_i^S \varphi)' \stackrel{def}{\equiv}_Q (K_i^E \varphi')^Q \qquad (L_i^S \varphi)' \stackrel{def}{\equiv}_Q (L_i^E \varphi')^Q$$

The proof of Prop. 23 required the presence of the awareness operator in $\mathcal{L}(K^S, A)$ (but not that of $A$ in $\mathcal{L}(K^E, A)$, given Prop. 21). As speculative knowledge treats unaware atoms as their most general consistent interpretation, there is no semantic difference (with respect to just speculative knowledge) between an agent being unaware of an atom and an agent (speculatively) knowing nothing about it.

The lower end of this expressivity hierarchy is also of theoretical interest but maybe less of practical interest. We have various other results, that are given here without proof. Clearly the propositional language $\mathcal{L}(\emptyset)$ is less expressive than all of $\mathcal{L}(K^E)$, $\mathcal{L}(L^E)$, $\mathcal{L}(\Box)$, and $\mathcal{L}(K^S)$. More interesting is that, although we already established that $\mathcal{L}(K^E)$ is equally expressive as $\mathcal{L}(K^E, A)$, still, $\mathcal{L}(\Box)$, $\mathcal{L}(L^E)$ and $\mathcal{L}(K^S)$ are strictly less expressive than, respectively, $\mathcal{L}(\Box, A)$, $\mathcal{L}(L^E, A)$ and $\mathcal{L}(K^S, A)$. Interestingly, $\mathcal{L}(\Box)$ and $\mathcal{L}(K^S)$ are incomparable. And so on ...

## 6. AXIOMATIZATION

In this section we present complete axiomatizations for our logics.

Table 1 presents an axiomatization $\mathbf{L}^\Box$ characterizing the validities of the language $\mathcal{L}^\Box$ in epistemic awareness models (the logic $\mathbf{L}^\Box$). This axiomatization is provided in [4], modulo a minor variation (see Section 8).

| All propositional tautologies | $A_i \top$ |
|---|---|
| $\top$ | $A_i \neg \varphi \leftrightarrow A_i \varphi$ |
| $\Box_i(\varphi \to \psi) \to (\Box_i \varphi \to \Box_i \psi)$ | $A_i(\varphi \wedge \psi) \leftrightarrow A_i \varphi \wedge A_i \psi$ |
| From $\varphi$ and $\varphi \to \psi$ infer $\psi$ | $A_i \Box_j \varphi \leftrightarrow A_i \varphi$ |
| From $\varphi$ infer $\Box_i \varphi$ | $A_i A_j \varphi \leftrightarrow A_i \varphi$ |

Table 1: Axiom system $\mathbf{L}^\Box$

**Theorem 24 (Soundness and completeness)**
Axiom system $\mathbf{L}^\Box$ is sound and complete for $\mathcal{L}^\Box$ with respect to epistemic awareness models.

PROOF. Soundness is proved by showing that axioms in $\mathbf{L}^\Box$ are valid and that its rules preserve validity. Completeness is proved by using the canonical model technique in the standard way.

Table 2 presents an axiomatization $\mathbf{L}^E$ characterizing the validities of the language $\mathcal{L}^E$ in epistemic awareness models. A similar axiomatization, but with a different completeness proof, was provided in [9]. See again Section 8 for further discussion.

| All propositional tautologies | $A_i \top$ |
|---|---|
| $\top$ | $A_i \neg \varphi \leftrightarrow A_i \varphi$ |
| $K_i^E(\varphi \to \psi) \to (K_i^E \varphi \to K_i^E \psi)$ | $A_i(\varphi \wedge \psi) \leftrightarrow A_i \varphi \wedge A_i \psi$ |
| $K_i^E \varphi \to A_i \varphi$ | $A_i K_j^E \varphi \leftrightarrow A_i \varphi$ |
| From $\varphi$ and $\varphi \to \psi$ infer $\psi$ | $A_i A_j \varphi \leftrightarrow A_i \varphi$ |
| From $\varphi$ infer $A_i \varphi \to K_i^E \varphi$ | |

Table 2: Axiom system $\mathbf{L}^E$

**Theorem 25 (Soundness and completeness)**
Axiom system $\mathbf{L}^E$ is sound and complete for $\mathcal{L}^E$ with respect to epistemic awareness models.

PROOF. Soundness is proved by showing that axioms in $\mathbf{L}^E$ are valid and that its rules preserve validity. Completeness is proved by using the canonical model technique in the standard way.

Table 3 presents an axiomatization $\mathbf{L}^S$ characterizing the validities of the language $\mathcal{L}^S$ in epistemic awareness models. In axiom * of Table 3, called **KS**, it is required that $p \notin v(\varphi)$.

| All propositional tautologies | $A_i \top$ |
|---|---|
| $\top$ | $A_i \neg \varphi \leftrightarrow A_i \varphi$ |
| $K_i^S(\varphi \to \psi) \to (K_i^S \varphi \to K_i^S \psi)$ | $A_i(\varphi \wedge \psi) \leftrightarrow A_i \varphi \wedge A_i \psi$ |
| $K_i^S \varphi \to (\neg A_i p \to K_i^S \varphi[p \backslash \psi])$ * | $A_i K_j^S \varphi \leftrightarrow A_i \varphi$ |
| From $\varphi$ and $\varphi \to \psi$ infer $\psi$ | $A_i A_j \varphi \leftrightarrow A_i \varphi$ |
| From $\varphi$ infer $K_i^S \varphi$ | |

Table 3: Axiom system $\mathbf{L}^S$

Since the axiomatization for the logic of speculative knowledge is novel, we provide the results in detail.

**Theorem 26 (Soundness)** *Every theorem of $\mathbf{L}^S$ is valid.*

PROOF. This is a quite standard proof, and we only need to examine the axioms and rules involving speculative knowledge. Axiom $K_i^S(\varphi \to \psi) \to (K_i^S \varphi \to K_i^S \psi)$ and the rule of necessitation for $K^S$ are straightforward, and are also found in [21].

Axiom **KS** is new. It says that if an agent speculatively knows a formula despite the formula using an atom of which the agent is unaware, then the agent would continue to know that formula if the atom were replaced with any other formula. This axiom captures the intuition of the speculative knowledge operator, where if an agent is unaware of an atom, the agent must assume the most general interpretation of that atom. In other words, this is according to the semantics for speculative knowledge.

To prove completeness, we use the canonical model technique.

**Definition 27 (Canonical model)** *The canonical model for $\mathsf{L}^S$ is a tuple $M^c = (S^c, R^c, \mathcal{A}^c, V^c)$ where*

- $S^c$ *is the set of all theories (maximal consistent sets) of $\mathsf{L}^S$;*
- *For any $i \in N$, $R_i^c$ is a binary relation on $S^c$ such that $R_i^c(\Phi, \Psi)$ iff for all $\varphi$, $K_i^S \varphi \in \Phi$ implies $\varphi \in \Psi$;*
- $\mathcal{A}^c : N \to S^c \to P$ *is such that $p \in \mathcal{A}_i^c(\Phi)$ iff $A_i p \in \Phi$;*
- $V^c : P \to S^c$ *is such that $\Phi \in V^c(p)$ iff $p \in \Phi$.*

**Lemma 28** *For all formulas $\varphi \in \mathcal{L}^S$ and all maximal consistent sets $\Phi \in S^c$, $v(\varphi) \subseteq \mathcal{A}_i^c(\Phi)$ iff $A_i \varphi \in \Phi$.*

PROOF. This follows directly from the definition of a canonical model (Definition 27) and the axioms involving awareness (right-hand side of Table 3).

**Lemma 29** *Suppose that $(M, s)$ is image-finite. Let $Th(M, s)$ be $\{\varphi \in \mathcal{L}^S \mid (M, s) \models \varphi\}$. Then $(M, s) \leftrightarroweq (M^c, Th(M, s))$.*

PROOF. This follows from the definitions of the canonical model (Def. 27) and awareness bisimulation (Def. 3). Define the relation $B = \{(s, Th(M, s)) \mid s \in \mathcal{D}(M)\}$. It can be easily seen that $B$ satisfies clauses **atoms** and **aware**. For **forth**, if $t \in R_i(s)$, then we have $(t, Th(M, t)) \in B$, and we note that $Th(M, t) \in R_i^c(Th(M, s))$ since if $(M, s) \models K_i^S(\varphi)$, then $(M, t) \models \varphi$. For **back**, if $\Phi \in R_i^c(Th(M, s))$, then for every formula $\varphi \in \Phi$ we must have $(M, s) \models L_i^S \varphi$. Since $M$ is image-finite there must be some $t \in R_i(s)$ such that for all $\varphi \in \Phi$, $(M, t) \models \varphi$. Therefore $(t, \Phi) \in B$ and we are done.

**Lemma 30 (Truth)** *For all formulas $\varphi \in \mathcal{L}^S$ and all maximal consistent sets $\Phi \in S^c$, $(M^c, \Phi) \models \varphi$ iff $\varphi \in \Phi$.*

PROOF SKETCH. This is shown by induction over the complexity of formulas; we only show the non-trivial case $K_i^S \varphi$.

Suppose $(M^c, \Phi) \models K_i^S \varphi$. Then for all $\Psi \in R_i^c(\Phi)$ and for all $(N, t) \leftrightarroweq_{\mathcal{A}_i^c(\Phi)} (M^c, \Psi)$ we have $(N, t) \models \varphi$. Suppose for contradiction that $K_i^S \varphi \notin \Phi$; then there must be some $\Psi \in R_i^c(\Phi)$ such that $\neg\varphi \in \Psi$ (this follows from the maximality of $\Phi$, from propositional reasoning, and the axioms for distribution of $K^S$ over $\to$ and necessitation for $K^S$). By induction hypothesis, $(M^c, \Psi) \models \neg\varphi$ and, since awareness bisimulation is reflexive, we have the required contradiction, so we must have $K_i^S \varphi \in \Phi$.

Now suppose that $K_i^S \varphi \in \Phi$ and define $Q := v(\varphi) \setminus \mathcal{A}_i^c(\Phi)$. From axiom **KS** we also have that

$$\left(K_i^S \varphi \wedge \bigwedge_{q \in Q} \neg A_i q\right) \to K_i^S \varphi[Q \setminus \overline{\psi}],$$

where $\overline{\psi}$ is any vector of formulas in one-to-one correspondence with $Q$ (so that $[Q \setminus \overline{\psi}]$ stands for simultaneous substitution). So $K_i^S \varphi[Q \setminus \overline{\psi}] \in \Phi$. Now suppose that $\Psi \in R_i^c(\Phi)$, and that $(N, t)$ is any model such that $(N, t) \leftrightarroweq_{\mathcal{A}_i(\Phi)} (M^c, \Psi)$.

By Theorem 26, $Th(N, t)$ must be a maximally consistent set. For every atom $p \in Q$, we define a characteristic formula, $\chi(p)$, that is true exactly when $p$ is in the sets reachable from $Th(N, t)$, up to the modal depth of $\varphi$. This can be done by taking the intersection of these sets with the closure set of $\varphi$ (all subformulas of $\varphi$ and their negations) and the set $\mathcal{A}_i^c(\Phi)$. Applying axiom **KS**, substituting $\chi(p)$ for $p$ we can show that $L_i^S \psi \in \Phi$ for all $\psi \subseteq \varphi$ where $(N, t) \models \psi$. Since $K_i^S \varphi \in \Phi$ for every substitution of $Q$, it follows that $(N, t) \models \varphi$ as required. Therefore, $(M^c, \Phi) \models K_i^S \varphi$ as required.

**Theorem 31 (Completeness)**
*Let $\Phi \subseteq \mathcal{L}^S$ and $\varphi \in \mathcal{L}^S$. Then $\Phi \models \varphi$ implies $\Phi \vdash \varphi$.*

# 7. DYNAMICS

## 7.1 Epistemic awareness action models

Epistemic awareness models represent the information of agents who may be uncertain about the truth of some propositional variables and unaware of others. The information of such agents can change via informational acts. Epistemic awareness action models represent awareness change and knowledge change. They were introduced in [22] for the logic of speculative knowledge. The definition adds a component for awareness to the action models of [1] (and a component for postconditions, as in [23]).

**Definition 32 (Epistemic awareness action model)**
*An epistemic awareness **action** model is a tuple $\mathsf{M} = (\mathsf{S}, \mathsf{R}, \mathsf{A}, \mathsf{pre}, \mathsf{post})$ where*

- $\mathsf{S}$ *is a non-empty set of actions;*
- $\mathsf{R} : N \to \mathcal{P}(\mathsf{S} \times \mathsf{S})$ *is an accessibility function;*
- $\mathsf{A} : \{+, -\} \to N \to \mathsf{S} \to \mathcal{P}(P)$ *is an awareness change function, indicating the disjoint sets of atoms each agent $i \in N$ will become aware (+) and unaware of (-) after the execution of $\mathsf{s} \in \mathsf{S}$;*
- $\mathsf{pre} : \mathsf{S} \to \mathcal{L}$ *is a precondition function specifying, for each action $\mathsf{s} \in \mathsf{S}$, the requirement for its execution;*
- $\mathsf{post} : \mathsf{S} \to P \to \mathcal{L}$ *is a postcondition function specifying, for each action in $\mathsf{s} \in \mathsf{S}$, how the truth value of each atomic proposition $p \in P$ will change.*

*A pair $(\mathsf{M}, \mathsf{s})$ with $\mathsf{s} \in \mathsf{S}$ is an epistemic awareness action.*

The language $\mathcal{L}$ of the preconditions and postconditions is a fixed parameter of this definition. We write $\mathsf{A}_i^+$ for $\mathsf{A}(+)(i)$ and $\mathsf{A}_i^-$ for $\mathsf{A}(-)(i)$.

**Example 4** *Particular kinds of epistemic awareness action models can be considered. Some examples:*

- *If $\mathsf{A}^+$ and $\mathsf{A}^-$ are both empty, the standard action models for knowledge change reappear.*
- *The singleton epistemic awareness action model with action $\mathsf{s}$ accessible to all agents, with trivial precondition and postcondition, and such that $\mathsf{A}_i^+(\mathsf{s}) = \{p\}$ for all agents $i$, represents 'all agents become aware of $p$' (without any knowledge change). For this action we write $\mathsf{A}^{+p}$. (Similarly, $\mathsf{A}^{-p}$, for becoming unaware of a variable.)*
- *The singleton epistemic awareness action model that is similar to the previous, but with precondition $\varphi$ and $\mathsf{A}_i^+(\mathsf{s}) = v(\varphi)$, represents a 'public announcement of a novel issue $\varphi$' — all agents become aware of the variables in $\varphi$ as part of the announcement. For this action we write $!_A \varphi$.*

We can now indicate how an epistemic awareness action model modifies an epistemic awareness model.

**Definition 33 (Action model execution)** *Let $M = (S, R, \mathcal{A}, V)$ be an epistemic awareness model, and let $\mathsf{M} = (\mathsf{S}, \mathsf{R}, \mathsf{A}, \mathsf{pre}, \mathsf{post})$ be an epistemic awareness action model. The epistemic awareness model $M \otimes \mathsf{M} = (S', R', \mathcal{A}', V')$ – the result of executing $\mathsf{M}$ in $M$ – is defined as follows:*

$$S' := \{(s, \mathsf{s}) \mid (M, s) \models \mathsf{pre}(\mathsf{s})\}$$
$$R'_i := \{((s, \mathsf{s}), (s', \mathsf{s}')) \mid s' \in R_i(s) \text{ and } (\mathsf{s}, \mathsf{s}') \in \mathsf{R}_i\}$$
$$\mathcal{A}'_i(s, \mathsf{s}) := (\mathcal{A}_i(s) \cup \mathsf{A}_i^+(\mathsf{s})) \setminus \mathsf{A}_i^-(\mathsf{s})$$
$$V'(p) := \{(s, \mathsf{s}) \mid (M, s) \models \mathsf{post}(\mathsf{s}, p)\}$$

The new set of states is the restricted cartesian product of $S$ and $\mathsf{S}$: a pair $(s, \mathsf{s})$ is a state in the new model iff $s$ satisfies $\mathsf{s}$'s precondition in $M$. Since the precondition is a formula of a language $\mathcal{L}$, we assume a satisfiability relation $\models$ that evaluates it. For the accessibility relation of the new model, we combine the accessibility relation of the 'static' model and the 'action' model: a state $(s', \mathsf{s}')$ is $R'_i$-accessible from state $(s, \mathsf{s})$ iff $s'$ is $R_i$-accessible from $s$, and $\mathsf{s}'$ is $\mathsf{R}_i$-accessible from $\mathsf{s}$. For the awareness function of each agent $i$ in each state $(s, \mathsf{s})$, we add to $\mathcal{A}_i(s)$ the atoms in $\mathsf{A}_i^+(\mathsf{s})$ and remove those in $\mathsf{A}_i^-(\mathsf{s})$ (in whatever order, as these sets are disjoint). For the valuation: an atomic proposition $p$ is true at state $(s, \mathsf{s})$ iff $s$ satisfies $\mathsf{post}(\mathsf{s}, p)$ in $M$.

## 7.2 Language and semantics

Instead of interpreting action models relative to a given logical language, we can also consider the set of action model frames as an additional parameter in an inductively defined language with a clause $[\mathsf{M}, \mathsf{s}]\varphi$ (where the precondition of actions should be lower in the inductive hierarchy); this stands for 'after execution of $(\mathsf{M}, \mathsf{s})$, $\varphi$ (is true)'.

**Definition 34 (Language)** *The language $\mathcal{L}(\otimes)$ extends any $\mathcal{L}$ with an additional inductive clause $[\mathsf{M}, \mathsf{s}]\varphi$, where $(\mathsf{M}, \mathsf{s})$ is an epistemic awareness action satisfying that: its domain is finite, the postcondition function changes the valuation of only a finite number of atomic propositions, and the awareness function returns two finite sets of atomic propositions. For $\mathcal{L}(\Box, A, \otimes)$ we write $\mathcal{L}^{\Box \otimes}$, for $\mathcal{L}(K^E, A, \otimes)$ we write $\mathcal{L}^{E\otimes}$, and $\mathcal{L}(K^S, A, \otimes)$ we write $\mathcal{L}^{S\otimes}$.*

**Definition 35 (Free variables)**
*An additional inductive clause $v([\mathsf{M}, \mathsf{s}]\varphi)$ is defined as*

$$v(\varphi) \cup \bigcup_{\mathsf{t} \in \mathcal{D}(\mathsf{M})} v(\mathsf{pre}(\mathsf{t})) \cup \bigcup_{\mathsf{t} \in \mathcal{D}(\mathsf{M}), p \in \mathcal{A}_i^+(\mathsf{t}) \cup \mathcal{A}_i^-(\mathsf{t})} (p \cup v(\mathsf{post}(\mathsf{t})(p)))$$

This definition of free variables formalizes that an agent is aware of an action $[\mathsf{M}, \mathsf{s}]$ if she is aware of all variables that occur in a precondition or postcondition of an action in the model $\mathsf{M}$. This can be called a conservative stance. For example, the agent can only be aware of $\Box_i p \to [\mathsf{A}^{+p}] K_i^E p$ (if the agent implicitly knows $\varphi$, then after becoming aware of $p$, the agent explicitly knows that $p$) if the agent is already aware of $p$ before the action. There is much wiggle room here that may also depend on philosophical considerations. For example, alternatively one could call a variable $p$ that occurs in a construct $[\mathsf{A}^{+p}] K_i^E p$ a *closed* variable. The variables that an agent is aware of now would then exclude those that she may become aware of later. We think this stance is conceptually problematic.

**Definition 36 (Semantics)**
*Let $M = (S, R, \mathcal{A}, V)$ and $s \in S$.*

$(M, s) \models [\mathsf{M}, \mathsf{s}]\varphi$ *iff* $(M, s) \models \mathsf{pre}(\mathsf{s}) \Rightarrow (M \otimes \mathsf{M}, (s, \mathsf{s})) \models \varphi$

*The set of validities of language $\mathcal{L}^{X\otimes}$ is called the logic $\mathsf{L}^{X\otimes}$ (for $X = \Box, E, S$).*

**Example 5** *The dynamic operator $[\mathsf{M}, \mathsf{s}]$ is not awareness bisimulation preserving. Consider this: The models $(M, s)$ and $(M', s)$ of the introduction are $p$ awareness bisimilar. And modally equivalent in $\mathcal{L}^E$. But after we make the agent aware of $p$, they are no longer $p$ awareness bisimilar. The formula $K_i^E K_i^E p$ is now a distinguishing formula. And therefore, $[\mathsf{A}^{+p}] K_i^E K_i^E p$ is true in $(M, s)$ and false in $(M', s)$. So $(M, s)$ and $(M', s)$ are not modally equivalent in $\mathcal{L}^{E\otimes}$.*

The dynamic operator $[\mathsf{M}, \mathsf{s}]$ is not awareness bisimulation preserving, but it is standard bisimulation preserving. The proof is similar for all three dynamic logics. In the proof we use modal equivalence in $\mathcal{L}^{X\otimes}$ (for $X = E, S, \Box$ of epistemic awareness states up to $Q$, denoted by $\equiv_Q^{X\otimes}$, defined analogously to $\equiv_Q^X$.

**Proposition 37** *Let $\varphi \in \mathcal{L}^{X\otimes}$, $Q \subseteq P$, and $(M, s), (M', s')$ given. If $(M, s) \simeq_Q (M', s')$, then $(M, s) \equiv_Q^{X\otimes} (M', s')$.*

PROOF. The proof is very similar to that in [22] for speculative knowledge. (Theorem 8 in [22] contains an error. It is here corrected.) The difference between implicit, speculative and explicit knowledge plays no role in the inductive case for action models. We only show that case.

**Inductive case $[\mathsf{M}, \mathsf{s}]\varphi$:** Suppose $(M, s) \models [\mathsf{M}, \mathsf{s}]\varphi$. Then $(M, s) \models \mathsf{pre}(\mathsf{s})$ implies $(M \otimes \mathsf{M}, (s, \mathsf{s})) \models \varphi$. By induction, $(M, s) \models \mathsf{pre}(\mathsf{s})$ iff $(M', s') \models \mathsf{pre}(\mathsf{s})$. The modal product construction in $(M \otimes \mathsf{M})$ is (standard) bisimulation preserving [1]; an easily observable fact when one realizes that pairs in the new accessibility relation require the first argument to be in the accessibility relation in the original model (given $(t, t') \in \mathfrak{R}[Q]$, the induced bisimulation $\mathfrak{R}'[Q]$ on the product is defined as $((t, \mathsf{t}), (t', \mathsf{t})) \in \mathfrak{R}'[Q]$). We now also have to satisfy the **aware** requirement. In the model $M \otimes \mathsf{M}$ the level of awareness $\mathcal{A}_i(t, \mathsf{t})$ is a function of the prior level of awareness $\mathcal{A}_i(t)$ in $t$ and the added or deleted propositional variables $\mathsf{A}_i^+(\mathsf{t})$ and $\mathsf{A}_i^-(\mathsf{t})$. As the prior awareness $\mathcal{A}_i(t)$ is the same in any $Q$ awareness bisimilar state $t'$, and the added or deleted atoms are also the same, the posterior awareness must therefore also be the same for any pairs $(t, \mathsf{t})$ and $(t', \mathsf{t})$ in the $Q$ awareness bisimulation. Therefore, $(M \otimes \mathsf{M}, (s, \mathsf{s})) \Leftrightarrow_Q (M' \otimes \mathsf{M}, (s', \mathsf{s}))$. Now using induction again, we conclude $(M' \otimes \mathsf{M}, (s', \mathsf{s})) \models \varphi$, and from that and $(M', s') \models \mathsf{pre}(\mathsf{s})$ we conclude $(M', s') \models [\mathsf{M}, \mathsf{s}]\varphi$.

Given the variety of knowledge and awareness changes that can be modelled by epistemic awareness action models, as shown in Example 4, the following is an important theorem. It demonstrates the adequacy of the framework.

**Theorem 38** *Let $(M, s)$ and $(M', s')$ be finite epistemic awareness states. Then there is an epistemic awareness action $(\mathsf{M}, \mathsf{s})$ such that $(M, s) \otimes (\mathsf{M}, \mathsf{s})$ is standardly bisimilar to $(M', s')$.*

PROOF. The proof is an extension of the one in [23]. We sketch the proof. *First*, delete the structure of $(M, s)$ by a public announcement of its characteristic formula (as $M$ is finite, this characteristic formula exists [2]). The result is a singleton epistemic awareness state consisting of $s$ only. It does not matter what the valuation is or the level of awareness because, *next*, we execute an epistemic awareness action with precondition true and with the exact structure of the target model $(M', s')$, using postconditions in actions instead of valuations in states (setting then the value of propositional variables to the value of the valuation in the corresponding state), and awareness change function in actions instead of awareness functions in states (setting then the level of awareness of propositional variables to that in the corresponding state). This last part on awareness is the extension with respect to [23].

An alternative construction is the straightforward execution in $(M, s)$ of an epistemic awareness action with the structure of the target model $(M', s')$, and then the result is an epistemic awareness state bisimilar to $(M', s')$ (but typically larger than in the previous construction, it now has size $|M \otimes M'|$ instead of size $|M'|$).

## 7.3 Axiomatization

We now give the axiomatization of the logic $\mathsf{L}^{\square\otimes}$. In Table 4 we only give the axioms involving action models. The ones for awareness after actions were presented in [22] and the one for implicit knowledge after action is novel, but has the standard shape of [1]. These are rewrite rules, that allow us to eliminate epistemic awareness action from formulas (given an innermost action model, one pushes it deeper and deeper into a formula until one of the first two axioms can be applied at which moment it has disappeared on the right-hand side). This proves the completeness of the axiomatization and the logic $\mathsf{L}^{\square\otimes}$ is therefore also equally expressive as the logic of implicit knowledge $\mathsf{L}^{\square}$.

---
$[\mathsf{M},\mathsf{s}]\top \leftrightarrow \top$
$[\mathsf{M},\mathsf{s}]p \leftrightarrow (\mathsf{pre}(\mathsf{s}) \to \mathsf{post}(\mathsf{s},p))$
$[\mathsf{M},\mathsf{s}]\neg\varphi \leftrightarrow (\mathsf{pre}(\mathsf{s}) \to \neg[\mathsf{M},\mathsf{s}]\varphi)$
$[\mathsf{M},\mathsf{s}](\varphi \wedge \psi) \leftrightarrow ([\mathsf{M},\mathsf{s}]\varphi \wedge [\mathsf{M},\mathsf{s}]\psi)$
$[\mathsf{M},\mathsf{s}]A_i\varphi \leftrightarrow \neg\mathsf{pre}(\mathsf{s})$  if $v(A_i\varphi) \cap \mathcal{A}_i^-(\mathsf{s}) \neq \emptyset$
$[\mathsf{M},\mathsf{s}]A_i\varphi \leftrightarrow (\mathsf{pre}(\mathsf{s}) \to A_i\varphi[\mathcal{A}_i^+(\mathsf{s})\backslash\top])$  otherwise
$[\mathsf{M},\mathsf{s}]\square_i\varphi \leftrightarrow (\mathsf{pre}(\mathsf{s}) \to \bigwedge_{\mathsf{t}\in\mathsf{R}_i(\mathsf{s})} \square_i[\mathsf{M},\mathsf{t}]\varphi)$
From $\varphi$ infer $[\mathsf{M},\mathsf{s}]\varphi$

---

Table 4: Axioms for action models in $\mathbf{L}^{\square\otimes}$

**Proposition 39** $\mathbf{L}^{\square\otimes}$ *is sound and complete.*

**Example 6** *To get an idea for the axioms involving awareness after actions, consider* $[\mathsf{A}^{+p}]A_ip$. *Surely we want the agents to be aware of p after becoming aware of p. The righthand side of this axiom computes to* $A_ip[p\backslash\top]$ *which is* $A_i\top$, *a theorem.*

*The other axiom applies when the agent becomes unaware. For example, consider* $[\mathsf{A}^{-p}]$, *which stands for 'the agents become unaware of* $\varphi$*' (not to be seen as gradual fading out, but as conscious abstraction). After that, the agents are no longer aware of p, so* $[\mathsf{A}^{-p}]A_ip$ *should be false. The righthand side of the axiom is* $\neg\mathsf{pre}([\mathsf{A}^{-p}])$. *The action* $[\mathsf{A}^{-p}]$ *is always executable: precondition* $\top$. *Its negation is therefore the contradiction* $\bot$, *as desired.*

## 7.4 Expressivity

In this short section we show that the logics $\mathsf{L}^{\square\otimes}$, $\mathsf{L}^{E\otimes}$, $\mathsf{L}^{S\otimes}$ are all equally expressive (and therefore, as $\mathsf{L}^{\square\otimes} = \mathsf{L}^{\square}$, all equally expressive as $\mathsf{L}^{\square}$).

**Proposition 40** $\mathsf{L}^{\square\otimes}$ *and* $\mathsf{L}^{E\otimes}$ *are equally expressive.*

PROOF. This follows from the following equivalences (embeddings). The first demonstrates that $\mathsf{L}^{\square\otimes} < \mathsf{L}^{E\otimes}$ and the second (wherein we use a familiar equivalence, but now within the language $\mathcal{L}^{\square\otimes}$ instead of $\mathcal{L}^{\square}$) that $\mathsf{L}^{\square\otimes} > \mathsf{L}^{E\otimes}$.

$$\begin{array}{rcl}\square_i\varphi & \Leftrightarrow & [\mathsf{A}^{+v(\varphi)}]K_i^E\varphi \\ K_i^E\varphi & \Leftrightarrow & A_i\varphi \wedge \square_i\varphi\end{array}$$

**Proposition 41** $\mathsf{L}^{E\otimes}$ *and* $\mathsf{L}^{S\otimes}$ *are equally expressive.*

PROOF. The same argument as in Prop. 23 applies here.

This is an unmistakable though somewhat (we think) surprising result. Even though the logic of implicit knowledge is *more* expressive than the logic of explicit knowledge, the dynamic logic of implicit knowledge is *equally* expressive as the dynamic logic of explicit knowledge. And similarly for speculative knowledge. Example 5 clearly demonstrates the increase of expressive power when dynamics are added: all of a sudden we can distinguish the models $(M, s)$ and $(M', s)$!

To conclude the picture — and this paper — the axiomatization for the dynamic logic of explicit knowledge is therefore simply the one wherein you write $K_i^E\varphi$ as $\square_i\varphi \wedge A_i\varphi$ and then derive that in $\mathbf{L}^{\square\otimes}$. This does not get us the axiomatization for the dynamic logic of speculative knowledge yet, a missing piece in this puzzle, but as the expressivity of this logic is now known, this seems of decidedly minor interest.

## 8. RELATED WORKS

Our epistemic awareness models are those of [4]. The language used there is $\mathcal{L}(\square, K^E, A)$, but it has the same expressivity as $\mathcal{L}(\square, A)$, since $K_i^E\varphi$ is definable as $\square_i\varphi \wedge A_i\varphi$ (see Proposition 20). The setting of [4] is otherwise different. They assume the accessibility relations to be serial, transitive and euclidean ($KD45$). For the axiomatization one can simply add the characterizing axioms. The complete axiomatization provided there defines awareness $A_ip$ by abbreviation as $K_i^E(p \vee \neg p)$.

Another pertinent investigation is [9]. It focusses on axiomatizations, not on expressivity issues. In [9], Halpern presents axiomatizations for the logics with languages $\mathcal{L}(\square, A)$, $\mathcal{L}(K^E, A)$ and $\mathcal{L}(K^E)$, for the model class where the ($KD45$) agents also know their own awareness: $t \in R_i(s)$ implies $\mathcal{A}_i(s) = \mathcal{A}_i(t)$. In the axiomatization for $\mathcal{L}(\square, A)$ we find this as $A_i\varphi \to \square_i A_i\varphi$ and $\neg A_i\varphi \to \square_i \neg A_i\varphi$. In the axiomatization for $\mathcal{L}(K^E, A)$ this property is, instead, described by an axiom $A_i\varphi \to K_i^E A_i\varphi$ and a rule IRR.: "If no propositional variables in $\varphi$ appear in $\psi$, then from $\neg A_i\varphi \to \psi$ infer $\psi$" (with the suggestion that the rule might be derivable from the axiomatization). The rule IRR. is also discussed in [10]. These additional features seem to explain that the completeness proof for the logic of explicit knowledge in [9] is more involved than ours.

The language $\mathcal{L}(K^E)$ is shown in [9] to have the same expressivity as $\mathcal{L}(K^E, A)$ but with the crucial difference that this is on models with euclidean accessibility relations and knowledge of awareness. In such models awareness can be defined in terms of explicit knowledge (as also done in [17]): $A\varphi \leftrightarrow K^E\varphi \vee K^E \neg K^E\varphi$. We recall that in our approach $A\varphi \leftrightarrow K^E(\varphi \vee \neg \varphi)$ (similar to [4], see above), but this equivalence does not hold on the more restricted model class.

Some recent studies on dynamics, such as [12, 8, 19] take a somewhat different approach to awareness, namely syntactic awareness, but employ similar ideas for the dynamics: updates of structures.

## 9. CONCLUSIONS AND FURTHER WORK

We have the described the logics of implicit, explicit, and speculative knowledge, related modal equivalence in these logics to different forms of bisimulation, compared their expressivity, and provided sound and complete axiomatizations. Then we investigated the dynamics of these logics, where we have shown that any conceivable change of knowledge or awareness can be modelled, we axiomatized the dynamic logic of implicit knowledge, and showed that all three dynamic logics are equally expressive.

Concerning further work, we wish to close some (we think) little gaps. The axiomatization of the logic of speculative knowledge with respect to $S5$ structures is not necessarily an extension of the current axiomatization. This is because the speculative knowledge operator has a built-in quantification over awareness bisimilar structures. Quantifying over structures in a more restricted model class therefore changes the semantics of speculative knowledge; and therefore, also its axiomatic properties. Another little gap is that, even though we know the expressivity of the dynamic logic of speculative knowledge, we do not have (as mentioned above) its axiomatization (with or without the $S5$ restriction).

Further ahead, there are alternative notions of knowledge beyond implicit / explicit / speculative that employ propositional awareness, for example: an agent knows a formula $\varphi$ in state $s$ iff in all accessible states $t$, $\varphi$ is true and the agent is aware of $\varphi$ (a version explored in [19]). Or consider knowledge employing a recursive version of awareness: agent $i$ is aware of $K_i^E \varphi$ in $s$ iff it is aware of $\varphi$ in $s$ and aware of $\varphi$ in all $t$ $i$-accessible from $s$. Alternative notions of knowledge would correspond to yet other notions of bisimulation.

The result of Theorem 38 that awareness action models can encode any form of knowledge and awareness change, is very strong. But from another perspective, it is also very weak, because typically only certain protocols or a given and commonly known set of actions are allowed. Investigating the dynamic logics of explicit and speculative knowledge for those settings may be relevant for game theory.

## 10. ACKNOWLEDGMENTS

We are in debt to the TARK reviewers whose very detailed and enthusiastic comments we have tried to do justice in the final version. We thank Joe Halpern for clarifying a technical detail concerning the relation between our axiomatization and that in his publication [9]. This work was done while Hans van Ditmarsch was employed by the University of Seville, Spain. Hans van Ditmarsch is also affiliated to IMSc, Chennai, as a research associate. Yì Wáng gratefully acknowledges funding support from the Major Project of National Social Science Foundation of China (No. 11&ZD088).